\documentstyle[preprint,aps,prd]{revtex}
\draft
\begin{document}

\title{Moment of Inertia and Superfluidity of a Trapped Bose Gas}

\author{S. Stringari}

\address{Dipartimento di Fisica, Universit\`a di Trento and INFM,
38050 Povo, Italy }

\date{ september 23, 1995 }

\maketitle

\begin{abstract}

The temperature dependence of the moment of inertia of a dilute Bose
gas confined in a harmonic trap is determined.
Deviations from the rigid value, due to the occurrence of Bose-Einstein
condensation, reveal the superfluid behaviour of the system.
In the noninteracting gas these deviations become  important at
temperatures of the order
of $T_cN^{-1/12}$. The role of interactions is
also discussed.

\end{abstract}


Evidence of Bose-Einstein condensation
in magnetically trapped gases of $^{87}Rb$ \cite{nist} and $^{7}Li$
\cite{rice} atoms has been recently reported.
These measurements are expected to open further stimulating
perspectives in the study of the phenomenon of Bose-Einstein
condensation (BEC) both from the experimental and theoretical point of view
\cite{BEC93}.
In this context
the understanding of the superfluid behaviour
of a Bose condensed atom cloud
is a question of primary importance that we plan to investigate
in the present letter.

A natural way to discuss superfluidity in a confined
system is to focus on its rotational properties \cite{baym}.
For a macroscopic system
the moment of inertia is given by the rigid value unless it exhibits
superfluidity. Crucial deviations from the rigid motion occur in
rotating liquid helium below the lambda temperature \cite{helium}.
Also finite systems, like some deformed atomic nuclei \cite{bm} and helium
clusters \cite{ceperley}, are known to exhibit important
superfluid effects in the moment of inertia.

The moment of inertia $\Theta$, relative to the $z$-axis, can be defined
as the linear response of the system to a rotational
field $H_{ext}=-\omega J_z$, according to the formula
\begin{eqnarray}
<J_z> = \omega \Theta
\label{deftheta}
\end{eqnarray}
where $J_z = \sum_i(x_ip^y_i-y_ip_i^x)$ is the $z$-component of the
the angular momentum and the average is taken on the state
perturbed by $H_{ext}$.
For a classical system the moment of inertia takes the rigid value
\begin{eqnarray}
\Theta_{rig} \equiv  mN<x^2+y^2>
\label{rigid}
\end{eqnarray}
where $N$ is the number of atoms in the trap.
Vice versa the quantum mechanical determination of $\Theta$ is
less trivial.
It involves a calculation of dynamical properties of the system
and, according to perturbation theory, can be written as
\begin{eqnarray}
\Theta = \frac{2}{Z}\sum_{n,m}e^{-E_m/k_BT}\frac{\mid<m\mid J_z\mid n>\mid^2}
{E_n-E_m}
\label{perturbation}
\end{eqnarray}
where $m$ ($n$) and $E_m$ ($E_n$) are eigenstates and eigenvalues
of the unperturbed hamiltonian and
$Z$ is the partition function. Generalization of (\ref{perturbation})
to the grand canonical
ensemble is straightforward.

In the following we calculate the moment of inertia of a dilute Bose
gas confined in a harmonic trap. We first consider the simplest
case of an ideal gas
described by the hamiltonian
\begin{eqnarray}
H = \sum_i \frac{{\bf p}^2_i}{2m} +
\sum_i\frac{m}{2}(\omega^2_xx^2_i + \omega^2_yy^2_i + \omega^2_zz^2_i) \,\, .
\label{oscillator}
\end{eqnarray}
The role of interparticle interactions will be discussed in the second part of
the work.
For the model hamiltonian (\ref{oscillator})
the moment of inertia is explicitly evaluated by solving the equation
\begin{eqnarray}
[H,X] = J_z
\label{defx}
\end{eqnarray}
for the operator $X$ which, according to (\ref{perturbation}),
determines the moment of inertia
through the relation $\Theta=<[J_z,X]>$. The explicit form of the operator
$X$ is found to be \cite{lippa}
\begin{eqnarray}
X = -\frac{i}{\hbar(\omega_x^2-\omega_y^2)}
\sum_i[( \omega_x^2+\omega_y^2)x_iy_i  +
\frac{2}{m} p_i^xp_i^y]
\label{resultforX}
\end{eqnarray}
and the moment of inertia takes the form
\begin{eqnarray}
\Theta = \frac{mN}{\omega^2_x-\omega^2_y} [(<y^2>-<x^2>)(\omega^2_x+\omega^2_y)
+2(\omega^2_y<y^2>-\omega^2_x <x^2>)]
\label{theta}
\end{eqnarray}
where the average is taken at  statistical equilibrium in the absence
of the perturbation $H_{ext}$ \cite{virial}. Notice that expression
(\ref{theta}) makes sense only if $\omega_x\ne\omega_y$.

Equation (\ref{theta}) is an  exact result for the
hamiltonian (\ref{oscillator}).
It applies to any value of $N$ and holds both for Bose and Fermi
\cite{nuclei} statistics.
For the explicit evaluation of the moment of inertia in a Bose gas
it is useful to evaluate the average radii $<x^2>$ and $<y^2>$ in the so called
macroscopic limit
where one separates the contribution arising
from the condensate from the one of the other excited states, whose
quantum numbers are approximated by continuous variables (semiclassical
approximation). This description is appropriate
for temperatures $k_BT$
much larger than the oscillator energies $\hbar\omega_x$,
$\hbar\omega_y$, $\hbar\omega_z$
and yields the following result for the depletion of the condensate \cite{kl}:
\begin{eqnarray}
N-N_0(T) = N\frac{T^3}{T_c^3}
\label{defN0}
\end{eqnarray}
with the critical temperature given by \cite{kl}
\begin{eqnarray}
k_BT_c = \hbar\omega_0(\frac{2}{Q(2)}N)^{1/3} \,\, .
\label{defTc}
\end{eqnarray}
Here $\omega_0^3=\omega_x\omega_y\omega_z$ and $Q(\eta)=\int_0^{\infty}
s^{\eta}/(\exp s-1)ds$. For the square
radius $<x^2>$ one finds the result
\begin{eqnarray}
N<x^2> = N_0(T) \frac{\hbar}{2m\omega_x} +
(N-N_0(T)) \frac{k_BT}{3m\omega^2_x}\frac{Q(3)}{Q(2)}
\label{x2}
\end{eqnarray}
and analogously for $<y^2>$. The first term of this equation gives
the contribution to $<x^2>$ arising from the particles in the condensate and
scales as $1/\omega_x$.
The second one is the contribution from the non condensed particles and
scales as  $1/\omega^2_x$. When inserted into (\ref{theta}),
the contributions to $<x^2>$ and $<y^2>$ which scale
as $1/\omega_x$ and $1/\omega_y$, give rise to the {\it
irrotational} value for the moment of inertia.
Conversely contributions which scale
as $1/\omega_x^2$ and $1/\omega_y^2$ yield
the {\it rigid} value. The final result is

\begin{eqnarray}
\Theta = \epsilon^2_0m<x^2+y^2>_0N_0(T) + m<x^2+y^2>_{nc}(N-N_0(T))
\label{theta12}
\end{eqnarray}
where the indices $<>_0$ and $<>_{nc}$ mean average on the Bose
condensed and non condensed components of the system. The quantity
\begin{eqnarray}
\epsilon_0 = \frac{<x^2-y^2>_0}{<x^2+y^2>_0}
\label{epsilon}
\end{eqnarray}
is the deformation parameter of the Bose condensed density.
In terms of the oscillator frequencies, it is given  by
$\epsilon_0=(\omega_y-\omega_x)
/(\omega_y+\omega_x)$.

The physics contained in equation (\ref{theta12}) is very clear.
The moment of inertia is the sum of two terms arising from
the atoms
in the condensate (superfluid component),
which contribute with their irrotational flow, and from
the particles out of
the condensate (normal component)
which rotate in a rigid way. These two distinct
contributions are at the origin of an interesting $T$-dependence of the
moment of inertia. In fact above $T_c$, where $N_0(T)=0$, the
moment of inertia takes
the classical rigid value (\ref{rigid}),
while at $T=0$, where all the atoms are in the Bose condensate,
is given by the irrotational value $\Theta_{irrot} =
\epsilon^2\Theta_{rig}$.
The explicit $T$ dependence is easily calculated
using the harmonic oscillator result
\begin{eqnarray}
\frac{<x^2+y^2>_{nc}}{<x^2+y^2>_0} =
\frac{2}{3}\frac{Q(3)}{Q(2)}
\frac{\omega^2_x+\omega^2_y}{\hbar(\omega_x+\omega_y)\omega_x\omega_y}k_BT
\,\, .
\label{ratio}
\end{eqnarray}

Assuming $\omega_x\simeq \omega_y \ne \omega_z$ and taking
the values $Q(2)=2.4$ and $Q(3)=6.5$,
one obtains the expression
\begin{eqnarray}
\frac{\Theta}{\Theta_{rig}} =
 1 - \frac{1-(T/T_c)^3}{1-(T/T_c)^3+1.7(T/T_c)^4(\lambda N)^{1/3}}
\label{thetafinal}
\end{eqnarray}
where $\lambda = \omega_z/\omega_x$.
Near $T_c$
the moment of inertia deviates very little from the rigid
value not only because of the smallness of the number of atoms
in the condensate, but also because
of the large $N^{1/3}$ factor in the denominator of Eq.(\ref{thetafinal}).
This is the consequence of the fact that the  radius
of the atoms in the condensate is much smaller than the one of the atoms
out of the condensate.
In order to observe {\it superfluid}
effects in the moment of inertia
one has to go to smaller temperatures of the order of
$T\simeq T_cN^{-1/12}$ where
the term proportional to $N^{1/3}$ in the denominator of Eq.(\ref{thetafinal})
is of the order of unity. It is worth noting that, even for large values of
$N$,
the relevant regime of temperatures is not much smaller
than $T_c$ and can still be large compared to the
oscillator temperature.
In the figure we report the prediction of Eq.(\ref{thetafinal}) for
two different values of $N$ and $\lambda=\sqrt8$ (this value characterizes
the magnetic trap of \cite{nist}).

Result (\ref{theta12}) for the moment of inertia
has been derived for an ideal Bose gas
confined in a harmonic trap. The same result is
expected to hold also for a weakly interacting gas
where the mean field description \cite{tony,siggia}
permits to obtain
coupled equations for the wave functions of the condensate and
of the single particle excited states. The
irrotational flow for the
condensate then follows from general properties of the equation for the phase.
On the other hand the rigid flow
for the particles out of the
condensate follows from the use of the semiclassical approximation
which should be applicable in a useful range of temperatures \cite{tony}.
Interactions can neverthelss
modify the value of the moment of inertia (\ref{theta12}) by changing
the temperature dependence of $N_0(T)$ as well as the value of the square
radii $<x^2+y^2>$ \cite{tony,siggia,oliva,ruprecht,bp}.

In order to give a first estimate of these effects
we have considered the Hartree-Fock equations at
temperatures such that the thermal depletion
of the condensate is small and one can  consequently ignore
the contribution of excited atoms to the self-consistent effective field.
The temperature should be however large enough in order to apply
the semiclassical approximation to the excited states. This regime
is the most interesting for the investigation
of superfluid effects in the moment of inertia.
Furthermore, in
order to obtain an  explicit estimate, we have considered the case
of strongly repulsive interactions where the
the kinetic term in the Gross-Pitaevskii equations for the
condensate  can be neglected \cite{tony,siggia}.

Under the above conditions (small thermal depletion,
semiclassical approximation for the excited states and strongly repulsive
interactions) the Hartree-Fock equations at finite temperature
yield the following simplified
expressions for the condensate ($\rho_0$) and non condensate
($\rho_{nc}$) densities:
\begin{eqnarray}
\rho_0({\bf r}) = \frac{m}{4\pi \hbar^2a}(\mu-v_{ext})
\theta(\mu-v_{ext})
\label{rho0}
\end{eqnarray}
and
\begin{eqnarray}
\rho_{nc}({\bf r}) = (\frac {mk_BT}{2\pi \hbar^2})^{3/2}g_{3/2}
(\exp(-\mid \mu - v_{ext}\mid/k_BT)) \,\, .
\label{rhonc}
\end{eqnarray}
In the above equations $a$ is the scattering length, $\theta$ is the unit-step
function, $g_{3/2}$ is the usual Bose function, $v_{ext}=
\frac{m}{2}(\omega^2_xx^2 + \omega^2_yy^2 + \omega^2_zz^2)$ is the harmonic
potential and
\begin{eqnarray}
\mu=\frac{1}{2} (15\sqrt{m}\omega_0^3N\hbar^2a)^{2/5}
\label{mu}
\end{eqnarray}
is the $T=0$ value of the chemical potential, fixed by the normalization
of $\rho_0$.
{}From Eqs.(\ref{rho0}) and (\ref{rhonc}) one gets the useful results (valid
if $(N-N_0)<<N$):
\begin{eqnarray}
\frac{N-N_0}{N} =
\frac{T^3}{T^3_c}F_0(\frac{\mu}{k_BT})
\label{depletion}
\end{eqnarray}
and
\begin{eqnarray}
\frac{<x^2+y^2>_{nc}}{<x^2+y^2>_0} = \frac{7F_2(\frac{\mu}{k_BT})}
{3F_0(\frac{\mu}{k_BT})}\frac{k_BT}{\mu}
\label{ratiobis}
\end{eqnarray}
with
$F_n(\mu/k_BT)=\int_0^{\infty}ds s^{2+n}g_{3/2}(exp(-\mid s^2-\mu/k_BT\mid)
/\int_0^{\infty}ds s^2g_{3/2}(exp(- s^2))$.

In Eq.(\ref{depletion}) $T_c$ is the critical temperature (\ref{defTc}) of the
noninteracting Bose gas.
We assume that this value is not significantly affected by the
interaction.

With respect to the predictions (\ref{defN0}) and (\ref{ratio})
of the ideal Bose gas, the interactions increase the thermal depletion
of the condensate
$(N-N_0(T))$ due to the (positive)
chemical potential entering Eq.(\ref{rhonc}).
On the other hand they can significantly decrease the ratio
$<x^2+y^2>_{nc}/<x^2+y^2>_0$ since the radius of the condensate density
is very sensitive to the repulsive forces.
As an example we have chosen $N=20000$ and the values
$a=100 a_0$ ($a_0$ is the Bohr radius),
$a_x =\sqrt{\frac{\hbar}{m\omega_x}} =1.65\times 10^{-4}cm\sim a_y$ and
$\lambda=
\sqrt{8}$, typical of the rubidium trapped gas of
Ref.\cite{nist}. With these values we find
$k_BT_c=36.1\hbar\omega_x$ and $\mu=11.6\hbar\omega_x$.
At $T=T_c/3$ the thermal depletion is increased from $0.04$ (prediction
of the noninteracting model) to $0.09$, the ratio
$<x^2+y^2>_{nc}/<x^2+y^2>_0$ is decreased from $21.7$ to $4.0$ and
the ratio $\Theta/\Theta_{rig}$ from $0.45$ to $0.3$,
thereby showing that in this case
the effects of superfluidity in the moment of inertia
are reinforced by the interactions.

A more systematic investigation
in a wider range of temperatures and
values of the scattering length (including the interesting
case of attractive forces)
requires the self-consistent solution of the Hartree-Fock equations
at finite temperature.
The present analysis, however, already permits to draw a first
important conclusion, i.e.  that superfluid effects in the moment of
inertia of a magnetically trapped Bose gas
should be observable at temperatures not dramatically smaller
than the critical temperature for Bose-Einstein condensation.

A final question concerns the onset of rotational instabilities
occurring at large frequencies $\omega$. For an axially symmetric gas
the stability is ensured below
the critical frequency
$\hbar\omega_c = min(\epsilon_{J_z}/J_z)$
where $\epsilon_{J_z}$ is the energy of a general excitation
carrying angular
momentum $J_z$. In the noninteracting case the value of $\omega_c$
coincides with the oscillator frequency $\omega_x$. In the presence of
interactions one should calculate the dispersion law of the
elementary excitations as well as the energy needed to
create
a vortex line \cite{bp}. Work in this direction is in progress.

Useful discussions with D. Brink, E. Cornell and F. Laloe are acknowledged.

\begin{figure}

\caption{Temperature dependence of the condensate fraction (full line)
and of the moment of inertia for $N=2000$ (dashed) and $N=20000$
(dot-dashed) atoms in the noninteracting gas (see the text).}

\end{figure}


\begin{thebibliography} {99}
\bibitem{virial} In order to derive result (\ref{theta}) we have used
     the identies $<p_x^2>=m^2\omega^2_x<x^2>$ and
$<p_y^2>=m^2\omega^2_y<y^2>$.
\end{thebibliography}
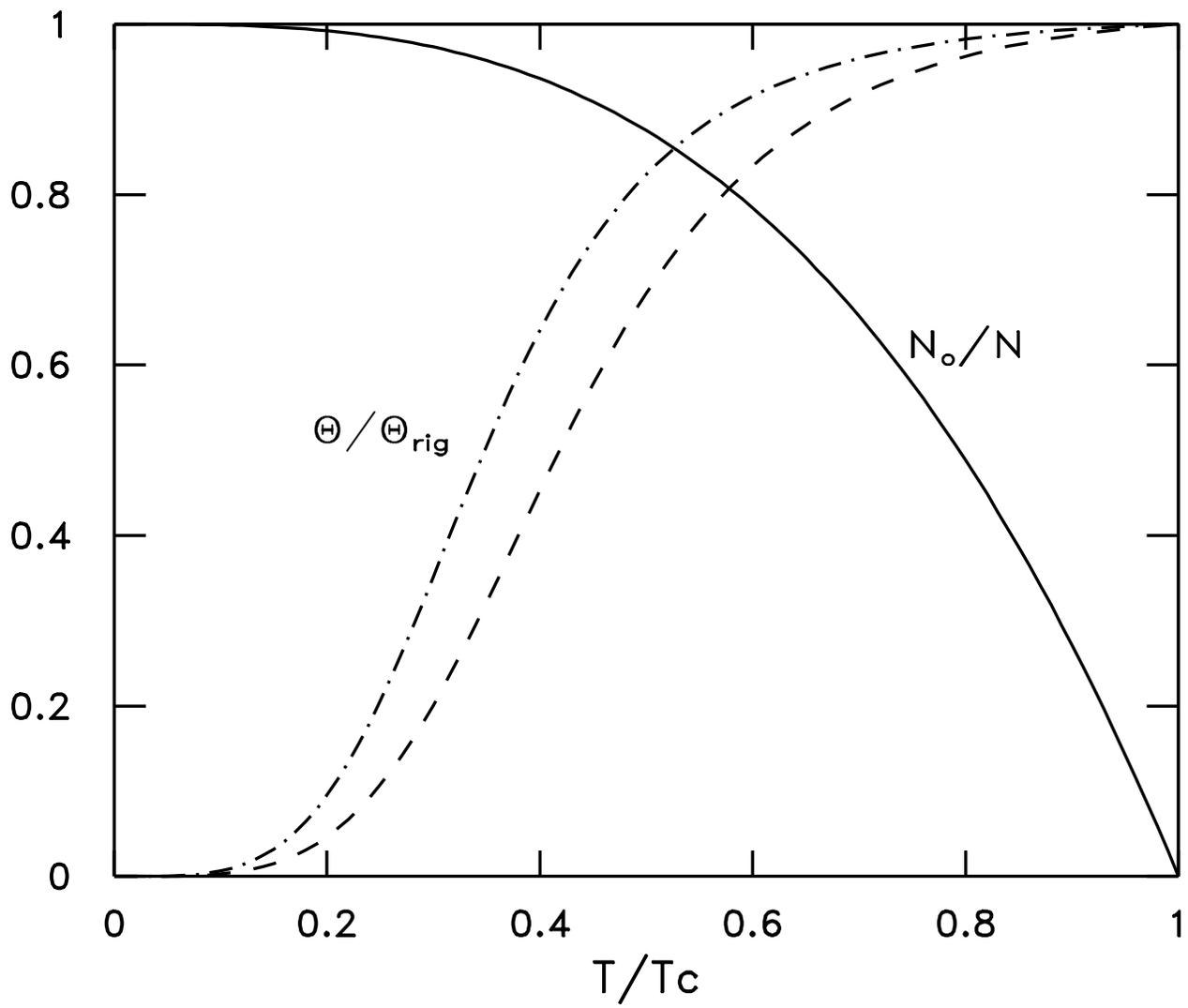
\end{document}